\documentclass[aps,prb,twocolumn,showpacs,superscriptaddress]
{revtex4-2}
\usepackage[utf8]{inputenc}
\usepackage{amsmath}
\usepackage[english]{babel}
\usepackage{graphicx, color}
\usepackage{hyperref}
\usepackage{natbib}
\usepackage{xcolor}

\definecolor{link}{rgb}{0.1,0.1,0.9}
\hypersetup{colorlinks=true,linkcolor=link,citecolor=link,urlcolor=link,linktocpage}
\usepackage{epstopdf}
\usepackage{color}
\usepackage{amsmath}
\usepackage{longtable}
\usepackage{amssymb}
\usepackage{siunitx}
\usepackage{amsfonts}
\usepackage{csquotes}

\begin{document}
	
\title{Observation of linear magnetoresistance and planar Hall effect in the Dirac semimetal PdTe }

\author{Bhawana Mehra}
\affiliation{ Quantum Materials and Devices Unit, Institute of Nano Science and Technology, Sector-81, Punjab, 140306, India.}
	
\author{ Bibek Ranjan Satapathy}
\affiliation{ Quantum Materials and Devices Unit, Institute of Nano Science and Technology, Sector-81, Punjab, 140306, India.}

\author{V S Abhishek}
\affiliation{ Department of Physical Sciences, Indian Institute of Science Education and Research Mohali, Sector 81, S. A. S. Nagar, Manauli, PO: 140306, India}

\author{Reena}
\affiliation{ Department of Physics, Indian Institute of Technology Ropar, Rupnagar, Punjab 140001, India.}

\author{Ritu Gupta}
\affiliation{ Department of Physics, Indian Institute of Technology Ropar, Rupnagar, Punjab 140001, India.}

 \author{Yogesh Singh}
\affiliation{ Department of Physical Sciences, Indian Institute of Science Education and Research Mohali, Sector 81, S. A. S. Nagar, Manauli, PO: 140306, India}

\author{S. Chakraverty}
\thanks{Corresponding author: suvankar.chakraverty@gmail.com}
\affiliation{ Quantum Materials and Devices Unit, Institute of Nano Science and Technology, Sector-81, Punjab, 140306, India.}

 \author{Amit Vashist}
\thanks{ Corresponding author: amitvashist42@gmail.com}
\affiliation{ Quantum Materials and Devices Unit, Institute of Nano Science and Technology, Sector-81, Punjab, 140306, India.}
\date{\today}

\begin{abstract}
 
PdTe is a Dirac semimetal that also exhibits superconductivity, providing an intriguing platform to explore topological superconductivity and unconventional magnetotransport phenomena. While the superconducting properties of PdTe have been extensively studied in recent years, the detailed magnetotransport phenomena have remained unexplored. Here, we present the first observation of linear magnetoresistance (LMR) and the planar Hall effect (PHE) in a high-quality single crystal of PdTe. We observe temperature-dependent unsaturated LMR in both in-plane ($B \parallel I$) and out-of-plane ($B \perp I$) configurations. The magnetoresistance (MR) shows a crossover from parabolic to linear dependence at the critical field $B_c$, and detailed analysis indicates that disorder-driven mobility fluctuations are the origin rather than the Abrikosov quantum-limit mechanism. Furthermore, prominent PHE has been observed by rotating the magnetic field within the plane of the sample. The detailed analysis of the field and temperature dependence of PHE-amplitude, along with the parametric plot, suggests that PHE originates predominantly from the anisotropic orbital magnetoresistance rather than the chiral anomaly. Our results demonstrate that although the Dirac point is close to the Fermi level, the observed LMR and PHE in the Dirac semimetal PdTe can be understood within a semiclassical transport framework, highlighting the importance of distinguishing between topological and conventional classical transport mechanisms in topological materials.

\end{abstract}

\maketitle	
	
\section*{1. Introduction}

The theoretical prediction and experimental realization of topological materials over the past decade have significantly broadened our understanding of rich electronic transport phenomena beyond the framework of conventional materials \cite{1,2,3}. Owing to their nontrivial topology and the presence of linear dispersion relations, these materials exhibit several exotic properties, such as extremely high MR and LMR, chiral-anomaly-induced negative MR, low effective mass and high carrier mobility, the anomalous Hall effect, and the PHE \cite{6,7,8,9,10}. Understanding these transport properties not only provides insights into fundamental physics but also offers opportunities to explore their use in future electronic and spintronic applications \cite{4,5}.

 Among these magnetotransport phenomena, LMR and PHE have been widely investigated in Dirac semimetals (DSMs). LMR has attracted significant attention because it defies the expectation of the quadratic field dependence of magnetoresistance, as predicted by the Lorentz force, in conventional metals and semiconductors. Large LMR has been observed in several materials, such as narrow-band-gap semiconductors, InSb, Ag/PtTe$_2$/W \cite{11,12}, topological insulators, Bi$_2$Te$_3$, Bi$_2$Se$_3$ \cite{5,13,14}, topological semimetals, CaCdSn, ZrGeSe \cite{6,15}, perovskite oxides, LaVO$_3$/KTaO$_3$ \cite{16,17}, and multilayer graphene \cite{18}. The physical origin of LMR in these materials is primarily explained either by the classical model proposed by  Parish and Littlewood \cite{19}  or by the quantum model proposed by Abrikosov \cite{20}. In the classical model, mobility fluctuations can result in LMR in disordered or inhomogeneous materials, whereas in the quantum model, LMR occurs when all charge carriers occupy the lowest Landau level, a regime known as the quantum limit. Additionally, PHE is another intriguing phenomenon that, unlike the conventional Hall effect, arises when a magnetic field is applied within the plane of the applied current and induces a transverse voltage \cite{21,22}. In topological semimetals, the observation of PHE was initially proposed as a signature of the chiral anomaly and associated Berry curvature and was therefore considered a useful indirect transport signature of the topological band structure  \cite{22,23}. However, theoretical calculations and experimental investigations have demonstrated that PHE in topological systems can also arise from other mechanisms, such as anisotropic orbital magnetoresistance, current jetting, and anisotropic magnetic scattering \cite{10,24,25,26}. Therefore, a careful and systematic investigation of both LMR and PHE is important for distinguishing between topological and conventional transport mechanisms.

 PdTe has recently been identified as a type-I Dirac semimetal, with the Dirac point lying close to the Fermi level \cite{27}. It is a non-layered material that crystallizes in a hexagonal crystal structure  \cite{28,30}. In addition to its topologically nontrivial electronic structure, PdTe also exhibits superconductivity, with a superconducting temperature ranging from 4.25 K to 4.6 K  \cite{34,31,32}. The nature of superconductivity in PdTe remains debated, with conflicting reports on whether it is a nodal or nodeless multigap superconductor \cite{31,59,33}. Its Dirac nature has been established using theoretical calculations, angle-resolved photoemission spectroscopy (ARPES), and quantum oscillation analysis  \cite{27,32,34}. Previous band-structure calculations show the presence of both electron and hole-type bands crossing the Fermi level \cite{34}, which is consistent with the observation of a nonlinear Hall signal in PdTe \cite{35,36}. Despite several reports on the superconducting nature of PdTe, its magnetotransport properties have not been well explored.

In this article, we present detailed angle-dependent and temperature-dependent magnetotransport properties of single-crystalline PdTe. We report, for the first time, the observation of LMR and PHE in the Dirac semimetal PdTe. A crossover from parabolic MR at low fields to LMR at high magnetic fields (up to 14 T) is observed for $B \parallel I$ and $B \perp I$. A two-band Hall model is employed to estimate the carrier concentrations and mobilities of charge carriers in PdTe at various temperatures. The relationship between the effective mobility and the crossover magnetic field suggests that classical mobility fluctuations induced by disorder may be a possible mechanism for the observed LMR. We also report PHE and anisotropic magnetoresistance (AMR) in PdTe. The field-dependent amplitudes (0–14 T) of both AMR and PHE deviate from quadratic behavior, as expected for the PHE associated with the chiral anomaly. The parametric plot shows a shock-wave like pattern, indicative of anisotropic orbital magnetoresistance.

\section*{2. RESULTS AND DISCUSSION}

\begin{figure}
        \centering
        \includegraphics[width=1\linewidth]{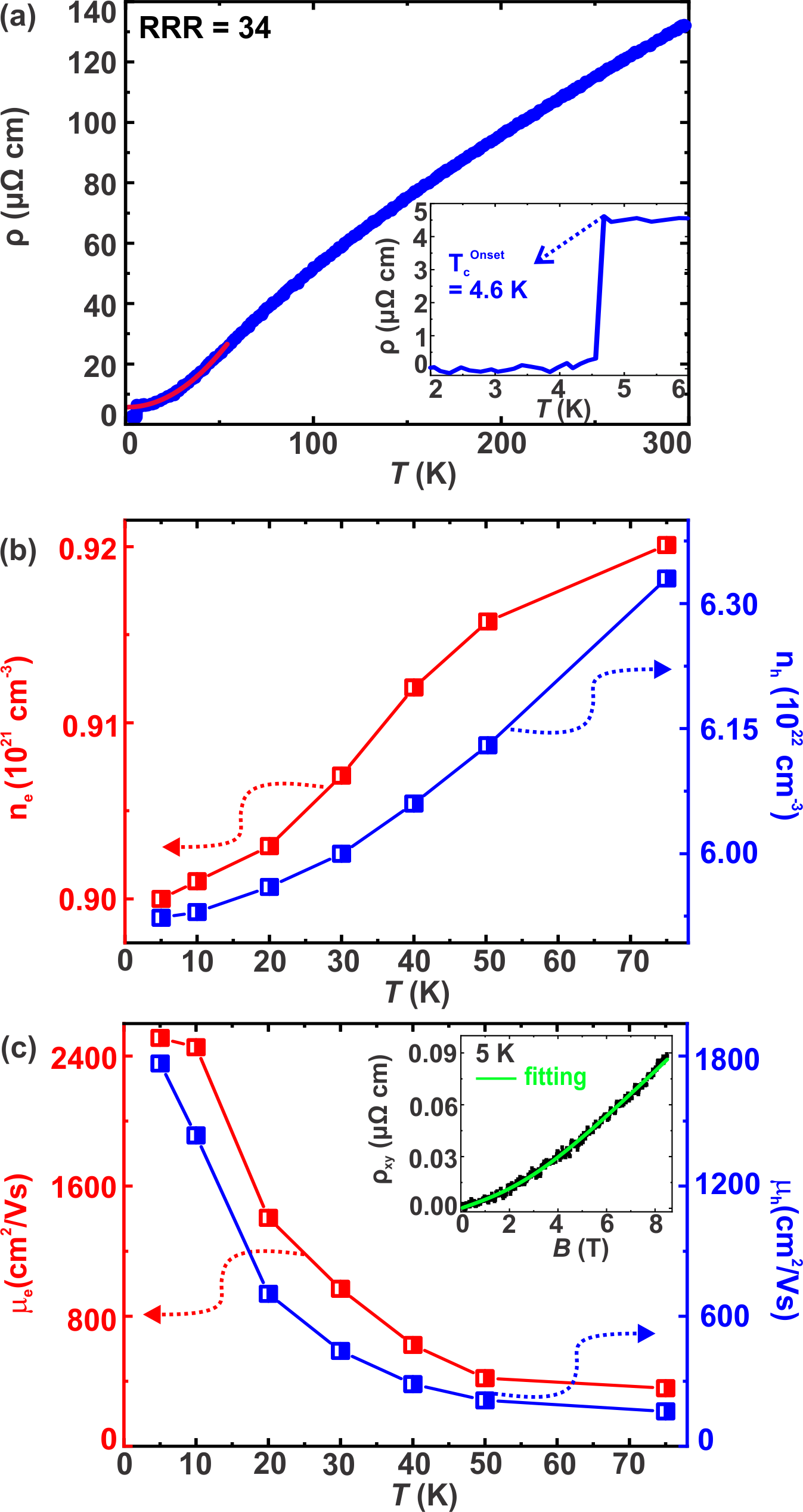}
        \caption{(a) Temperature dependence of the resistivity $\rho(T)$ from 2 to 300 K of the PdTe single crystal at zero magnetic field, showing metallic behaviour with a residual resistivity ratio of 34. The red curve is a fit to the low-temperature data from the temperature range 0-53 K. The inset highlights the superconducting transition with an onset $T_c$ = 4.6 K. (b) Temperature dependence of the electron ($n_e$, red, left axis) and hole ($n_h$, blue, right axis) carrier concentrations extracted from Hall-effect measurements between 5 and 75 K, both increasing monotonically with temperature. (c) Corresponding temperature dependence of the electron ($\mu_e$, red, left axis) and hole ($\mu_h$, blue, right axis) mobilities. The inset represents the Hall resistivity $\rho_{xy}(B)$ as a function of applied magnetic field B at 5 K (black squares), together with a two-band model fit (green line) used to extract the carrier densities and mobilities as shown in (b) and (c).}
        \label{Fig-1}
    \end{figure}

High-quality single-crystalline samples of PdTe were prepared using a melt-growth technique. The stoichiometric amounts of Pd and Te were heated to 1000 °C in an evacuated quartz tube and then slowly cooled. The detailed temperature profile is provided in a previous report \cite{32}. The chemical composition and phase purity were confirmed using energy-dispersive X-ray spectroscopy (EDS) and powder X-ray diffraction (PXRD), respectively \cite{32}. Electrical transport measurements were performed using a Quantum Design Physical Property Measurement System (QD-PPMS Dynacool, 14 T). Figure.~\ref{Fig-1}(a) shows the temperature dependence of the resistivity measured from 300 to 2 K. The data exhibit the typical metallic behaviour of PdTe, with the resistivity decreasing upon cooling.  Additionally, a superconducting transition has been observed at around 4.6 K, as shown in the inset of figure~\ref{Fig-1}(a). The residual resistivity ratio (RRR), calculated as $\rho(300)/\rho_{0} $,  is estimated to be approximately 34, indicating the high quality of the measured crystal. The $\rho_{0} $ was calculated by fitting the low-temperature data (5 - 53 K) using the expression $\rho$ = $\rho_{0}$ + AT$^2$. The RRR value is found to be higher than those reported in several previous studies  \cite{27,33,36}. The antisymmetrized Hall resistivity $\rho_{xy}$ as a function of magnetic field at 5 K is shown in the inset of figure~\ref{Fig-1}(c). The field dependence of the Hall resistivity exhibits clear nonlinear behavior, indicating the presence of more than one type of charge carrier, consistent with previous theoretical band-structure calculations \cite{34}. To determine carrier concentrations and mobilities, a two-band Hall model was employed. The solid green curve represents the two-band Hall-model fit to the experimental data at 5 K (inset of figure~\ref{Fig-1}(c)). The fitting yields the electron carrier density, $n_e$ = 0.9 × 10$^{21}$ $cm^{-3}$, hole carrier density, $n_h$ = 5.9 × 10$^{22}$ $cm^{-3}$, electron mobility, $\mu_e$ = 2512 $cm^2/Vs$, and hole mobility, $\mu_h$ = 
1766 $cm^2/Vs$ at 5 K, consistent with previous reports \cite{35,36}. The Hall effect was measured at various temperatures and a detailed analysis using the two-band Hall model is presented in the Supporting Information (Figure S1). The temperature dependence of electron and hole carrier concentrations and mobilities is shown in figures~\ref{Fig-1}(b) and ~\ref{Fig-1}(c), respectively. As expected, the carrier concentrations increase, whereas carrier mobilities decrease with increasing temperature for both electron and hole carriers.

\begin{figure*}
    \centering
    \includegraphics[width= 1\linewidth]{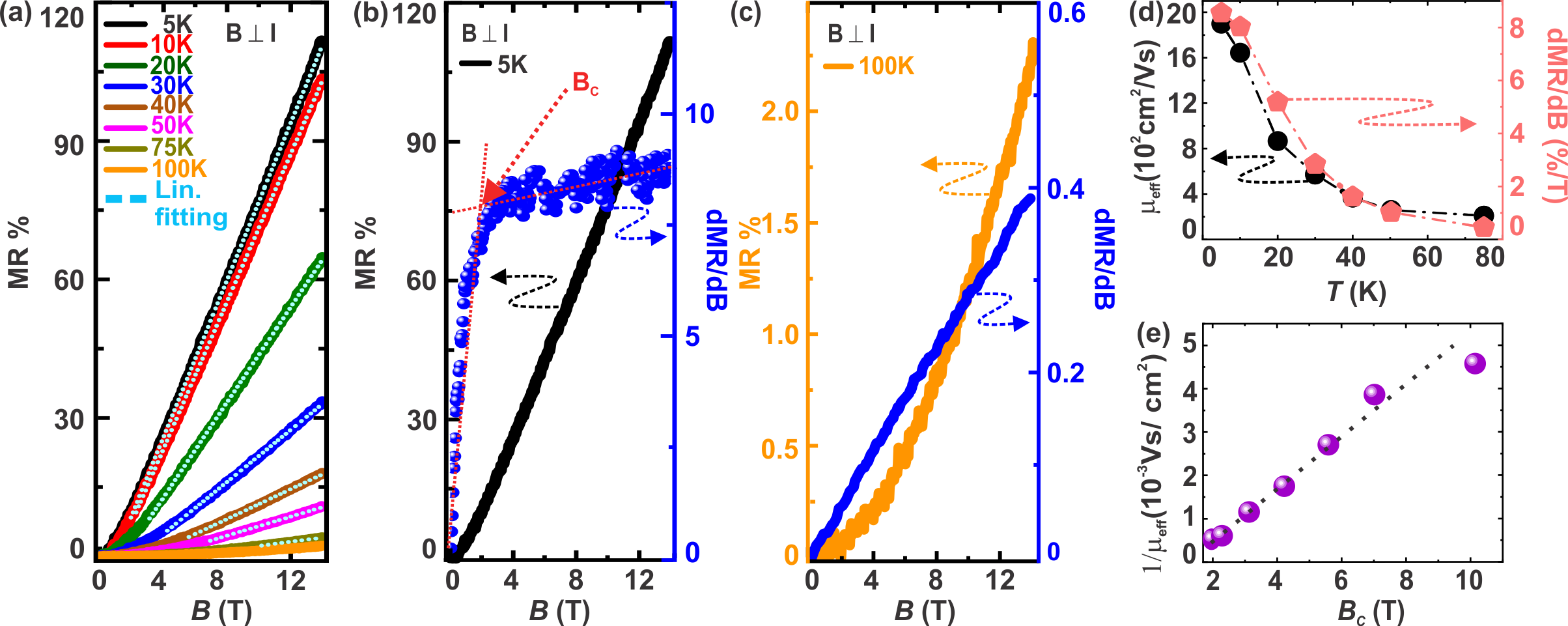}
    \caption{(a) Field dependence \% magnetoresistance (MR\%) has been plotted for different temperatures from 5 to 100 K in out-of-plane orientations. The dashed cyan lines show linear fits to the high-field region. (b) MR\% (black, left axis) and its field derivative dMR/dB (blue, right axis) at 5 K. The derivative shows a sharp initial rise followed by saturation, with the intersection of the red dotted lines identifying the characteristic crossover (critical) field $B_c$, marking the transition from quadratic to linear MR. (c) Corresponding MR\% (orange, left axis) and dMR/dB (blue, right axis) at 100 K, where the derivative varies smoothly with field without a clear saturation feature, in contrast to the high-field  behavior in (b). (d) Temperature dependence of the effective mobility $\mu_{eff}$ (black circles, left axis), and dMR/dB (pink pentagons, right axis), both of which decrease systematically with increasing temperature. (e) Inverse effective mobility (1/ $\mu_{eff}$) versus crossover field $B_c$ extracted at each temperature, showing an approximately linear relationship (black dotted line). }
    \label{Fig-2}
\end{figure*}

To further elucidate the magnetotransport properties of PdTe, we next investigate the longitudinal resistivity as a function of applied magnetic field at various temperatures. Figure~\ref{Fig-2}(a) displays the MR \% = {$(\rho{(B)}-{\rho(0))}/\rho{(0)}$} × 100, as a function of magnetic field at temperatures ranging from 5 to 100 K for the $B \perp I$ configuration. The MR increases with decreasing temperature, reaching 112\% at 5 K and 14 T for $B \perp I$, whereas 75\% for $B \parallel I$ configuration (Supporting Information, Figure S2). The observed MR in PdTe is anisotropic and is higher than that reported for some topological semimetals such as, NbTe$_2$, SrAgSb, Ag$_{0.05}$PdTe$_2$, and CaAgBi \cite{37,38,39,40} while being lower than that of WTe$_2$, IrTe$_2$ \cite{41,42}. For $B \parallel I$, the MR remains positive over the entire magnetic-field range at all measured temperatures, as shown in the Supporting Information (Figure S2). This behaviour suggests the absence of a negative longitudinal magnetoresistance associated with the chiral anomaly. The MR exhibits a conventional quadratic $B^2$-dependence at low fields and a linear $B$-dependence at high fields. This crossover from quadratic to linear MR is characterized by a critical field ($B_c$). The $B_c$ is estimated from the first derivative of the MR (intersection of a vertical and horizontal line) as shown in figure~\ref{Fig-2}(b). Initially, the derivative varies linearly with magnetic field at low fields, consistent with a $B^2$-dependence of the MR. As the magnetic field increases, the derivative approaches a constant value, indicating linear magnetoresistance at high fields. This crossover from $B^2$ to $B$ gradually becomes less pronounced as temperature increases and disappears near 100 K, as shown in figure~\ref{Fig-2}(c). At this temperature, the MR exhibits an approximately quadratic $B^2$-dependence over the entire measured field range, as indicated by the linear dependence of dMR/dB on magnetic field (blue curve in figure~\ref{Fig-2}(c)). A similar crossover from quadratic MR at low fields to LMR at high fields is also observed for $B \parallel I$) , as shown in the Supporting Information (Figure S2. (a)-(c)). The Kohler plot MR vs $B/\rho_0$  is shown in Supporting Information (Figure S2.(d)), where $\Delta \rho$ is the change in resistance with field $B$ and $\rho_0$ is zero field resistivity at particular temperatures. According to Kohler's rule, for a single dominant carrier type, the normalized MR should be a universal function of $B/\rho_0$ and the data obtained at different temperatures should collapse onto a single curve \cite{61}. As shown in Supporting Information (Figure S2(d)), the MR curves do not collapse onto a single curve, indicating a deviation from Kohler's scaling. This deviation suggests that multicarrier transport contributes to the magnetotransport in PdTe, consistent with the nonlinear Hall effect observed in PdTe.

We now discuss the possible origin of LMR in the Dirac semimetal PdTe. The origin of LMR can be described within two widely used theoretical frameworks: the classical random-resistor-network model proposed by Parish and Littlewood \cite{19} and the quantum model proposed by Abrikosov \cite{20}. PdTe exhibits a crossover from quadratic to linear MR, similar to that observed in several Dirac semimetals, such as SrMnBi$_2$ and LaAgSb$_2$ \cite{44,45}. Where LMR has been discussed in the context of the quantum model. In the quantum model, LMR is expected when the system reaches the quantum limit, where the charge carriers are confined to the lowest Landau level. The magnetic field required to reach the quantum limit depends on the carrier density of a material and can be estimated using the quantum-limit criterion given by 
$B_c$ = $\hbar (3\pi^2n)^{2/3}/{2e}$ \cite{46}. For PdTe, the carrier concentration is of the order of 10$^{28}$ $m^{-3}$. Considering the lowest carrier density, n = 0.9 x 10$^{21}$ $cm^{-3}$, obtained from the Hall data (Figure~\ref{Fig-1}), the magnetic field required to reach the quantum limit is estimated to be approximately 2.49 kT. This field is much larger than the experimentally accessible field range and is therefore inconsistent with the observed LMR. Furthermore, the condition for the occupation of only the lowest Landau level is $k_{B}T<<\hbar\omega_c$. Considering the experimentally determined $B_c$  at different temperatures and an effective mass of approximately, 0.35 $m_0$ \cite{32}, this condition is satisfied only at 5 K. However, LMR is observed up to approximately 75 K. Furthermore, we did not observe the quantum oscillations in the MR data up to 14 T at the lowest measured temperature, indicating the absence of Landau quantization. These results suggest that the observed LMR in PdTe cannot be explained by the Abrikosov quantum-limit mechanism.

The classical model proposed by Parish and Littlewood, which is based on disorder-induced mobility fluctuations in the material, is another mechanism used to explain LMR \cite{19,60}. In this model, the inhomogeneity in material can lead to fluctuation in the local charge carrier mobility, resulting in LMR. This model has been invoked to explain LMR in several materials, such as silver chalcogenides \cite{48}, as well as in topological materials like WTe$_{2}$, Bi$_{2}$Se$_{3}$ \cite{14,43,49}. According to the Parish–Littlewood model, the LMR is governed by two parameters: the effective mobility ($\mu_{eff}$) and the width of the mobility distribution $\Delta \mu$.  In the weak-disorder regime, $B_c$ is expected to scale inversely proportional to effective mobility {$(B_c \propto {\mu_{eff}^{-1})}$} , As discussed below, our system obeys $(B_c \propto {\mu_{eff}^{-1})}$, suggesting that PdTe lies in the weak-disorder regime.This interpretation is consistent with the high crystalline quality of our crystal, as indicated by high RRR value. Figure~\ref{Fig-2}(d) displays the temperature dependence of the $\mu_{eff}$ and the slope of the high-field LMR, dMR/dB. The effective mobility is calculated using $\mu_{eff}= \frac{n_{e}{\mu_{e}}^{2}+n_{h}{\mu_{h}}^{2}}{n_{e}\mu_{e}+n_{h}\mu_{h}}$, while dMR/dB is obtained from a linear fit to the MR in the high-field region beyond $B_c$. This figure shows that both the $\mu_{eff}$  and the high-field MR slope exhibit similar temperature dependence, suggesting a close relationship between the carrier mobility and the magnitude of LMR. This behavior is consistent with the classical Parish–Littlewood mechanism \cite{19,60}. Furthermore, the most compelling evidence for a disorder-induced LMR is that critical field should scale linearly proportional to the inverse of effective mobility (i.e $(B_c \propto {\mu_{eff}^{-1})}$). We therefore plot 1/$\mu_{eff}$ as a function of $B_c$ at corresponding temperatures, as shown in figure~\ref{Fig-2}(e). A linear relationship is observed between $B_c$ and 1/$\mu_{eff}$ supporting the proposed classical mechanism. Therefore, based on the above analysis, we conclude that LMR in PdTe is most likely attributed to the classical Parish–Littlewood  model. 

\begin{figure*}
    \centering
    \includegraphics[width=0.95\linewidth]{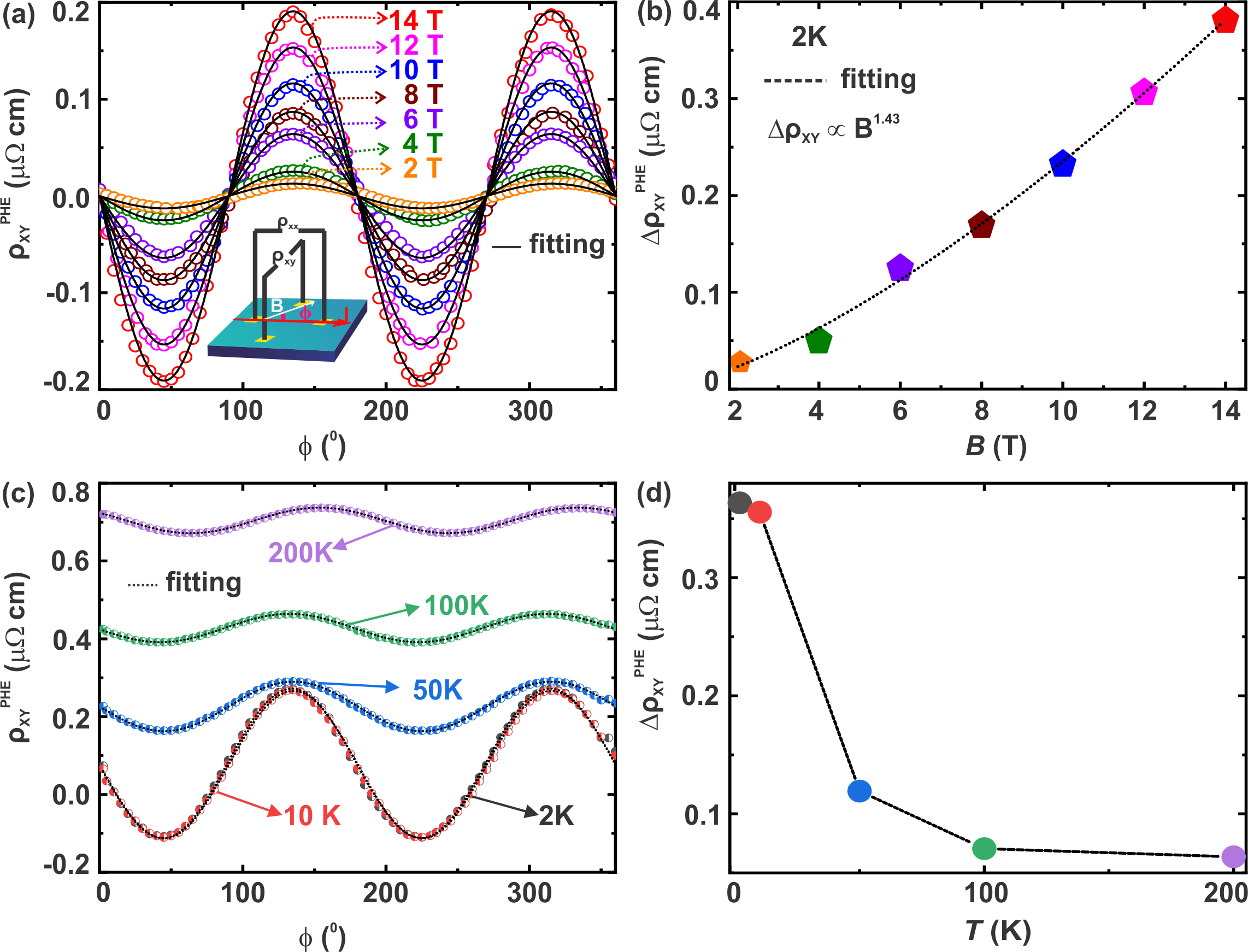}
    \caption{(a) Planner Hall resistivity $\rho_{xy}^{PHE}$ as a function of in-plane field angle $\phi$ at 2 K, for fields from 2 to 14 T (colored circles), with solid black lines showing fits to the expected sinusoidal angular dependence. Inset shows schematic of the measurement geometry, showing the field angle $\phi$ relative to the current direction I. (b) Field dependence of $\Delta\rho_{xy}^{PHE}$ (colored symbols), extracted by fitting the curves in figure~\ref{Fig-1}(a), using eq.~\ref{eq-1}, and black dotted line through the data represents the power law fit. (c) Angular dependence of $\rho_{xy}^{PHE}$ at 14 T, measured at various temperatures with black dotted lines showing fits to the data. (d) Temperature dependence of the extracted amplitude $\Delta\rho_{xy}^{PHE}$ at 14 T.} 
    \label{fig-3}
\end{figure*}

\begin{figure*}
    \centering
    \includegraphics[width=0.95\linewidth]{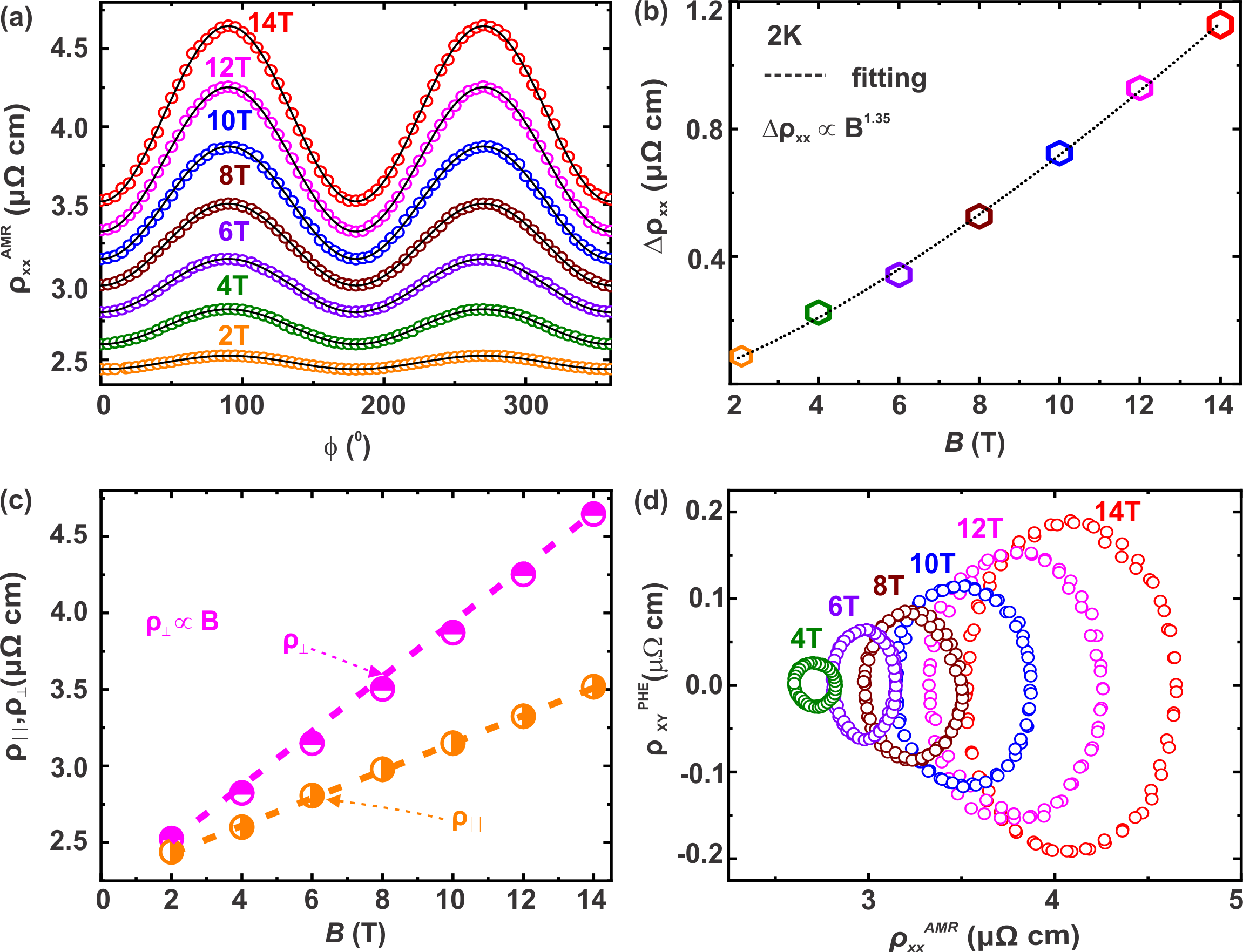}
    \caption{(a) Angular dependence of the anisotropic magnetoresistance (AMR) resistivity, ($\rho_{xx}^{AMR}$), measured at 2 K under magnetic fields ranging from 2 to 14 T. The solid curves represent fits to the experimental data. (b) Field dependence of the AMR amplitude $\Delta\rho_{xx}$ (colored hexagons), with black dotted line through the data represents the power law fit. (c) Field dependence of  $\rho_\parallel$ (orange) and $\rho_\perp$ (magenta). The dashed lines are eye guided lines, showing that $\rho_\perp$ scales linearly with B ($\rho_\perp$ $\propto B)$ and grows faster than $\rho_\parallel$. (d) Parametric plots of the planar Hall resistivity ($\rho_{xy}^{PHE}$)  as a function of $\rho_{xx}^{AMR}$, traced by sweeping $\phi$ through a full 360$^\circ$ rotation at each fixed field (4–14 T). The resulting closed loops, whose size increases systematically with field, reflect the shared angular periodicity and relative phase offset between the AMR and PHE signals, indicating an underlying anisotropic orbital mechanism.}
    \label{fig-4}
\end{figure*}

We next investigate the angle-dependent magnetotransport properties of the PdTe single crystals through PHE and anisotropic magnetoresistance (AMR). To study the in-plane magnetotransport properties, the crystal is rotated within the x-y plane under different magnetic fields. The magnetic field and current are kept parallel to the crystal plane, and the angle $\phi$ is defined as the angle between the magnetic field and the current direction, as shown in the insert of figure~\ref{fig-3}(a). The transverse and longitudinal resistivity measured in this configuration are referred to as the planar Hall resistivity ($\rho_{xy}^{PHE}$) and AMR($\rho_{xx}^{AMR}$). Figure~\ref{fig-3}(a) displays the angular dependence of $\rho_{xy}^{PHE}$ at various applied magnetic fields at 2 K. The PHE exhibits a characteristic 180$^\circ$ periodicity. The measured signal shows valleys near 45$^\circ$ and 225$^\circ$, while the peaks occur near 135$^\circ$ and 315 $^\circ$, consistent with several previous reports on PHE in topological semimetals such as NiTe$_2$, GdPtBi, MoTe$_2$,VAl$_3$, Cd$_3$As$_2$ \cite{23,25,50,51,52}.  It is important to mention that, the measured transverse signal may include contributions from ordinary Hall due to a small misalignment of the magnetic field with respect to the sample plane. The ordinary Hall signal is odd with respect to magnetic-field reversal, whereas the PHE signal is even. Therefore, measurements are performed under both positive and negative magnetic fields, and the PHE contribution shown in all relevant figures are obtained by averaging the two measurements. Another possible contribution arises from asymmetry in the Hall electrodes. If the Hall electrodes are not perfectly aligned along the transverse direction, a part of the longitudinal resistivity can be detected as a transverse voltage. This produces an additional contribution with a $cos^{2}\phi$ angular dependence. Since ${\textit{cos}}^{2}(\pi -\phi ) = cos^{2}\phi$ the longitudinal contribution can be separated from the PHE by comparing the signals measured at $\phi$ and $\pi-\phi$. A further contribution may arise from the sample's non-uniform thickness, which provides a nearly angle-independent geometrical background \cite{22}. We considered all these factors and plotted the angular dependence of PHE as shown in figure~\ref{fig-3}(a). The observation of PHE in topological materials has been frequently associated with the chiral anomaly \cite{51,53}, however, in addition to the chiral anomaly, several other mechanisms can contribute to the observed PHE, such as anisotropic magnetic scattering \cite{54}, anisotropic orbital magnetoresistance \cite{10}, and topological surface states \cite{55}. Since PdTe is a non-magnetic material, a magnetic-scattering origin of the PHE is unlikely. Furthermore, the relatively high carrier concentration suggests that the surface states of PdTe are likely strongly coupled to or submerged within the bulk states \cite{27,34}. Therefore, the observed PHE may have a significant contribution from bulk transport mechanisms, particularly anisotropic orbital magnetoresistance. To investigate the origin of PHE in PdTe, the intrinsic $\rho_{xy}^{PHE}$ is fitted by a theoretically derived semiclassical Boltzmann expression given as \cite{21,22}:

        \begin{equation}
	 	\rho_{xy}^{PHE} = -\Delta\rho \sin\phi\cos\phi  
          \label{eq-1}
	    \end{equation}

 where ${\Delta\rho = \rho_\parallel-\rho _\perp }$. Here, $\rho _\parallel $ represents the resistivity when the magnetic field is parallel to the current ($\phi= 0^\circ$), while $\rho _\perp$ represents the resistivity when the magnetic field is perpendicular to the current ($\phi= 90^\circ$). The quantity $\Delta\rho$, therefore, represents the difference between the resistivities for the two field orientations and determines the amplitude of the PHE. The solid curve through the data at various magnetic fields in figure~\ref{fig-3}(a) represents the fitting using eq.~\ref{eq-1}.  The extracted values $\Delta\rho^{PHE}$ as a function of magnetic field are shown in figure~\ref{fig-3}(b). The field dependence of the amplitude can be fitted by a power law, $\varDelta {\rho_{xy}^{PHE}}\ \propto \ B^{m}$ with an experimentally obtained exponent of approximately m = 1.4 as shown in figure~\ref{fig-3}(b). When the PHE originates from the chiral anomaly, the field dependence of its amplitude $\Delta\rho_{xy}^{PHE}$ is expected to show quadratic dependence on the magnetic field \cite{22}. In the present PdTe system, however, the experimentally observed field dependence of $\Delta\rho_{xy}^{PHE}(B)$ deviates from this expected behaviour. Therefore, the observed PHE cannot be attributed solely to the chiral anomaly. It is worth mentioning that similar exponential behavior with an exponent close to 1.4 has previously been reported in topological materials such as NiTe$_2$, Ni$_3$Bi$_2$Se$_2$, Co$_3$In$_2$S$_2$ and  PdTe$_2$ \cite{10,25,56,57}. Furthermore, figure.~\ref{fig-3}(c) shows the angular dependence of the PHE at different temperatures under a magnetic field of 14 T. The experimental data were fitted using eq.~\ref{eq-1}, with the corresponding fitting is shown as black solid lines. The PHE amplitude $\Delta\rho_{xy}^{PHE}$ extracted from these fits is plotted as a function of temperature in figure.~\ref{fig-3}(d). The PHE amplitude decreases sharply with increasing temperature up to approximately 50 K, after which the rate of decline slows at higher temperatures. This indicates that the PHE signal persists even at elevated temperatures. In $Cd_{3}As_{2}$ , where the PHE has been attributed to the chiral anomaly \cite{50}, the PHE amplitude shows only a small variation with increasing temperature in the low-temperature region, while significant changes are observed at higher  temperatures.

 To further investigate the origin of the PHE, we measured the angular dependence of the in-plane longitudinal resistivity at 2 K across different magnetic fields, as shown in Fig. 4(a). The resistivity exhibits a peak and dip at 90$^\circ$ and 180$^\circ$ for $B \parallel I$. The AMR data were fitted using the following equation \cite{21,22}:

\begin{equation}
	 	\rho_{xx} = \rho_\perp-\Delta\rho \cos^2\phi  
         \label{eq-2}
	    \end{equation}

The corresponding fitted curves are shown as black solid lines in figure~\ref{fig-4}(a). As the magnetic field increases, the overall resistivity increases gradually. From the fitting, the values of $\Delta\rho$, and $\rho_\perp$, $\rho_\parallel$ were extracted, as shown in figure.~\ref{fig-4}(b) and ~\ref{fig-4}(c), respectively.
The field dependence of the AMR amplitude follows a power law, $\varDelta {\rho_{xy}}\ \propto \ B^{1.34}$ ,which is close to the field dependence obtained from the PHE ($\varDelta \rho_{xy}\ \propto \ B^{1.43}$ , figure~\ref{fig-3}(b)). The similar power-law behaviour of the AMR and PHE amplitudes supports the consistency of the measurements and their analysis, suggesting a common origin for the PHE and AMR in PdTe. In the case of a chiral-anomaly-induced PHE, $\rho_\perp$ is generally expected to be approximately independent of the magnetic field, while $\rho_\parallel$ should decrease with increasing magnetic field, as observed in compounds such as WTe$_2$ \cite{53}. However, as shown in the figure.~\ref{fig-4}(c), $\rho_\perp$  increases approximately linearly with magnetic field, while $\rho_\parallel$ also increases, although at a relatively slower rate. The increase of both $\rho_\perp$  and $\rho_\parallel$ with magnetic field differs from the behaviour expected for a dominant chiral-anomaly contribution. Instead, it suggests orbital magnetoresistance arising from anisotropic transport parameters, such as the effective mass, scattering time, and carrier mobility. 

For further analysis, we plotted the planar Hall resistivity $\rho_{xy}^{PHE}$  against the longitudinal resistivity $\rho_{xx}^{AMR}$  at different magnetic fields. This type of representation is known as a parametric plot, where the angle $\phi$ is used as the parameter. As shown in figure~\ref{fig-4}(d), the parametric curves expand in a prolate-like form towards larger values of $\rho_{xx}$ as the magnetic field increases, without showing saturation up to 14 T. In systems where the PHE is dominated by the chiral anomaly, such as Na$_3$Bi and GdPtBi, the corresponding parametric curves form approximately concentric circles that expand uniformly with increasing magnetic field \cite{58}. In contrast, for anisotropic orbital magnetoresistance with the magnetic field applied in the plane, the parametric curves start from relatively small values of $\rho_{xx} $ and progressively expand towards larger $\rho_{xx}$ as the magnetic field increases, producing a characteristic shock-wave-like pattern. The shock-wave-like pattern observed in our measurements, therefore, supports an anisotropic orbital-magnetoresistance origin of the PHE. Although the PHE angular dependence can be well fitted using the equation associated with the chiral-anomaly model, the deviation of the field dependence from the expected $B^{2}$ dependence behaviour, the absence of negative longitudinal magnetoresistance for $B \parallel I$ (Figure S2, supporting information), and the shock-wave-like parametric pattern indicate that the chiral anomaly is unlikely to be the dominant origin of the observed PHE. Instead, the results suggest that anisotropic magnetoresistance, arising from anisotropic orbital transport, plays a significant role in the PHE observed in the PdTe single crystal.

\section*{3. CONCLUSIONS}

In summary, we have systematically investigated the magnetotransport properties of a single crystal of PdTe through the Hall effect, magnetoresistance, and angular-dependent magnetotransport measurements. The Hall effect is nonlinear, providing evidence of electron–hole multiband transport and consistent with the violation of Kohler’s rule. The magnetoresistance is found to be positive for both $B \perp  I$ and $B \parallel I$ measurement configurations, suggesting the absence of a chiral anomaly. The MR evolves from a quadratic $B^2$-dependence at low magnetic fields to a linear B-dependence at higher fields. The linear relationship between the inverse of effective mobility and the critical field (${\mu_{eff}}^{-1}\propto B_c$), together with the systematic temperature dependence of the carrier mobility and its correlation with the slope of the LMR, is consistent with the mobility-fluctuation mechanism proposed by Parish and Littlewood as the origin of LMR. The angular-dependent magnetotransport measurements reveal the observation of both the PHE and AMR in PdTe. The field dependence of $\Delta\rho$, extracted independently from the PHE and AMR measurements, exhibits a similar behavior, suggesting a common underlying transport mechanism. Furthermore, both ${\rho }_{\parallel }$ and ${\rho }_{\perp} $ remain positive and increase with increasing magnetic field, providing further evidence against the presence of a chiral anomaly. Additionally, the parametric plot exhibits a shock-wave-like feature, consistent with anisotropic orbital magnetotransport. Overall, our angular-dependent measurements suggest that the observed PHE in PdTe originates predominantly from anisotropic orbital magnetoresistance rather than from a chiral anomaly. Our results highlight that, despite the topological non-trivial band structure of PdTe, which hosts a Dirac point located near the Fermi level, its magnetotransport properties, including LMR and PHE, can be understood within the framework of classical transport mechanisms.

\medskip

\begin{center}
\textbf{Acknowledgments}
\end{center}
A.V. acknowledges the Department of Science and Technology (DST), India, for financial support through the INSPIRE Faculty Fellowship (Faculty Reg. No. IFA21-PH 279) and the Institute of Nano Science and Technology (INST), Mohali, for experimental facilities. B.M. acknowledges DST for financial support through the project DST/INSPIRE Fellowship/2022/IF220590. RG acknowledges support from the Anusandhan National Research Foundation (ANRF) under the PMECRG scheme (Grant No.ANRF/ECRG/2024/003930/PMS).

\medskip

\begin{center}
\textbf{Supporting Information}
\end{center}
Supporting Information is available from the author.

\medskip
\begin{center}
\textbf{Conflict of Interest}
\end{center}
The authors declare no conflict of interest.

\bibliographystyle{apsrev4-2}
\bibliography{mybib}

\end{document}